\documentclass{elsart}

\usepackage{amsmath,amssymb,epsfig}
\usepackage{bbm}

\begin{document}


\newcommand{\be}{\begin{equation}}
\newcommand{\ee}{\end{equation}}

\newcommand{\bea}{\begin{eqnarray}}
\newcommand{\eea}{\end{eqnarray}}

\newcommand{\Reals}{\mathbb{R}}     
\newcommand{\Com}{\mathbb{C}}       
\newcommand{\Nat}{\mathbb{N}}       

\newcommand{\id}{\mathbb{I}}       

\newcommand{\Real}{\mathop{\mathrm{Re}}}
\newcommand{\Imag}{\mathop{\mathrm{Im}}}
\def\O{\mbox{$\mathcal{O}$}}
\def\sgn{\text{sgn}}

\begin{frontmatter}


\title{Hopf bifurcations in time-delay systems with band-limited feedback}

\author{Lucas Illing\corauthref{cor}} and
\corauth[cor]{Corresponding author.}
\ead{illing@phy.duke.edu}
\author{Daniel J. Gauthier}

\address{Department of Physics and Center for Nonlinear and Complex
 Systems, Duke University, Durham, North Carolina, 27708, USA}


\begin{abstract}
 We investigate the steady-state solution and it's bifurcations in time-delay systems with band-limited feedback. This is a first step in a rigorous study concerning the effects of AC-coupled components in nonlinear devices with time-delayed feedback.
We show that the steady state is globally stable for small feedback gain and that local stability is lost, generically, through a Hopf bifurcation for larger feedback gain.
We provide simple criteria that determine whether the Hopf bifurcation is supercritical or subcritical based on the knowledge of the first three terms in the
Taylor-expansion of the nonlinearity.
 Furthermore, the presence of double-Hopf bifurcations of the steady state is shown, which indicates possible quasiperiodic and chaotic dynamics in these systems.
As a result of this investigation, we find that AC-coupling introduces fundamental differences to systems of Ikeda-type [Ikeda et al., Physica D 29 (1987) 223-235] already at the level of steady-state bifurcations, e.g. bifurcations exist in which limit cycles  are created with periods other than the fundamental ``period-2'' mode found in Ikeda-type systems.
\end{abstract}

\end{frontmatter}

\section{Introduction}

Recently there has been considerable interest in chaotic devices operating  in the radio-frequency (RF) regime for applications such as ranging and communication. As an example, such devices are studied as a possible signal source for radar applications because chaotic waveforms have the desirable properties of a  large frequency-bandwidth and a fast decay of correlations \cite{Lukin,Corron}. Furthermore, microwave \cite{Mykolaitis,Ott_Chaos}, optoelectronic \cite{illing,goedgebuer-qe,goedgebuer-qe2004,laspaper}, and optic \cite{VanWiggeren} devices are considered for communication systems since they can generate chaos with frequencies that match the frequency range of the communication infrastructure and provide advantages such as increased privacy \cite{goedgebuer-qe2004} and high power efficiency \cite{Ott_Chaos}.

 Many RF-devices are most accurately modeled by delay differential equations (DDEs) because the time it takes for signals to propagate through the device components is comparable to the time scale of the dynamics.
Furthermore, many of the chaotic devices are designed explicitly to include a time-delayed feedback \cite{Mykolaitis,Ott_Chaos,illing,goedgebuer-qe,goedgebuer-qe2004,laspaper,VanWiggeren}  because it introduces advantageous features such as the ability to tune the complexity of the dynamics by adjusting the delay \cite{Farmer} and the possibility of control in the presence of substantial control loop latency \cite{blakely-prl}.
In modeling, it is important to take the frequency characteristics of the feedback into account.
 
At high speed, many components are AC-coupled, which means that low-frequency signals below the frequency cut-off are suppressed. As a consequence, the time-delayed feedback signal is band-pass filtered because in addition to the cut-off at low frequencies, high frequencies are suppressed due to the finite response time of device components. 
Thus, DDEs describing band-limited feedback are the appropriate model for many of the devices that have been developed.

In the literature, we find cases where a model for the device-dynamics is constructed which ignores high-pass filtering in the feedback, but the experimental device has AC-coupled components. See for example, Ref.~\cite{Mykolaitis}. This suggest that the effect of AC-coupling on the dynamics is considered to be ignorable. However, counter-examples exist, such as the work of Goedgebuer \emph{et al.} \cite{goedgebuer-cs} and Blakely \emph{et al.} \cite{laspaper}.
Goedgebuer \emph{et al.} find that the inclusion of a high-pass filter in the feedback drastically increases the dimensionality of the chaotic attractor when compared to the same system  without such filter \cite{goedgebuer-cs}. 
Similarly, in our work on a high-speed chaotic time-delay system \cite{laspaper}, we found that it was necessary to model the effects of AC-coupling in order to explain the observed bifurcation of the steady state and the route to chaos.
These counter-examples demonstrate that high-pass filtering due to AC-coupled components can influence the dynamics substantially, indicating that a better theoretical understanding of the effects of high-pass filters in time-delay systems is needed.
As a first step in this direction, this paper provides  a detailed mathematical analysis of the bifurcations from the steady-state solution of systems with band-limited feedback.

\begin{figure}
\centerline{\includegraphics[width=0.5 \columnwidth]{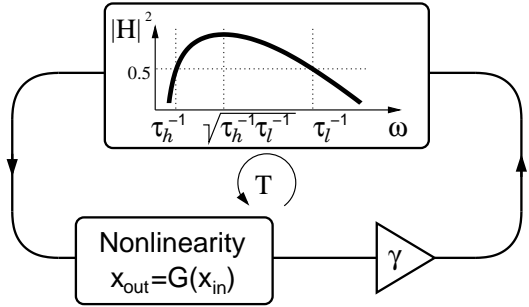} }
\caption{Sketch of a time-delay system with passive nonlinearity $G$, amplifier of gain $\gamma$, and band-limited delayed feedback.}
\label{fig:setup}
\end{figure}

The essential building blocks of the class of delay systems studied in this paper include a passive nonlinear element, a feedback loop, and a way to provide amplification (see  Fig.~\ref{fig:setup}).
 The nonlinear element maps the input to the output, $x_{out}=G(x_{in})$, the feedback loop is responsible for the delay $T$, and the control parameter $\gamma$ is a measure of the signal amplification.
Taking into account the band-pass characteristics of the feedback results in models that are two-dimensional DDEs in the simplest case, where, for simplicity, it is assumed the transfer characteristics of the feedback can be approximated by a  two-pole band-pass filter.
\begin{equation}
\frac{d z_1(\tilde{t})}{d\tilde{t}} = - \frac{z_1(\tilde{t})}{\tau_h} + \frac{d z_2}{d\tilde{t}}(\tilde{t}), \qquad
\tau_l \frac{d z_2(\tilde{t})}{d\tilde{t}}= - z_2(\tilde{t}) + \gamma~G\left[z_1(\tilde{t}-T)\right].
\label{model-urform}
\end{equation}
Here, the time-scale $\tau_h$ is related to the corner frequency of the high-pass filter through $\omega_h=\tau_h^{-1}$ and $\tau_l$ is related to the corner frequency of the low-pass filter through $\omega_l=\tau_l^{-1}$.

In this paper we study both the global and local stability of the steady-state solution of (\ref{model-urform}) for general nonlinear functions $G$. We derive explicit equations for the Hopf-bifurcation curves and use center-manifold techniques to provide simple criteria for the type of these bifurcations.
 Furthermore, we show that double-Hopf bifurcations of the steady state are possible, which indicates possible quasiperiodic and chaotic dynamics in these systems.

To the best of our knowledge, Hopf bifurcations of (\ref{model-urform}) have not been studied in detail.
On the other hand, the scalar DDE that result by taking the limit where the corner frequency of the high-pass filter goes to zero has been studied intensely both experimentally  \cite{Gibbs,ohtsubo,goedgebuer-pre} and theoretically \cite{ikeda-optcomm,Nardone,ikeda-physicaD,Hale_zamp,Zapp,Nizette,Erneux}, starting in 1979 with the pioneering work of Ikeda \cite{ikeda-optcomm}. 
In the discussion section we will therefore compare delay systems with band-limited feedback and Ikeda-type scalar DDEs to pinpoint the new dynamic features that arise due to the presence of the high-pass filter.

%
%

The paper is organized as follows. In Sec.~\ref{sec:model} we present the model equations.
In Sec.~\ref{sec:globstab}, a sufficient condition for global stability of  the steady-state solution is given. 
In Sec.~\ref{sec:locstab}, equations determining the boundary of local stability are derived and it is shown that the steady state becomes generically unstable via a Hopf bifurcation. The main focus of this section is the solution of the relevant characteristic equation. 
In Sec.~\ref{sec:normal_form}, center-manifold theory is used to determine whether the Hopf bifurcation is subcritical or supercritical. 
In Sec.~\ref{sec:examples}, three numerical examples are discussed. We conclude with a discussion section, Sec.~\ref{sec:discussion}.

\section{The Model}
\label{sec:model}

To simplify our analysis, we bring model (\ref{model-urform}) in a more convenient form by introducing variables $t,x,$ and $y$ through
\be
t = \tilde{t} \left(\frac{1}{\tau_l}+\frac{1}{\tau_h}\right), \qquad
x=z_1, \qquad y=\frac{\tau_h}{\tau_h+\tau_l}(z_1-z_2+\gamma G[0]),
\ee
to obtain
\begin{equation}
\begin{split}
\frac{d x(t)}{d t} &= - x(t) + y(t) + \gamma f[x(t-\tau)], \\
\frac{d y(t)}{d t} &= - r x(t).
\end{split}
\label{model}
\end{equation}
Here, the dimensionless delay is $\tau=T \; (\tau_l^{-1}+\tau_h^{-1})$, $r=\tau_l \tau_h (\tau_l+\tau_h)^{-2}$, and  $f[x(t-\tau)]= \tau_h (\tau_h+\tau_l)^{-1} (G[x(t-\tau)]-G[0])$. The nonlinear function $f$ is defined such that $f(0)=0$. 

There are three dimensionless parameters that influence the dynamics: the gain $\gamma$, the  strictly positive delay $\tau$, and $r$ ($0<r \le 1/4$).
The parameter $r$ is related to the angular frequency at which the transfer-function of the bandpass-filter is maximum.  Indeed, the frequency that maximizes transmission is $\omega_{max}=\sqrt{\omega_l \omega_h}$ (see  Fig.~\ref{fig:setup}), which, in the new coordinates, corresponds to the dimensionless angular frequency
\be
\Omega_{max}= \frac{\omega_{max}}{\omega_l + \omega_h } = \sqrt{r}.
\ee 
Additionally, $r$ provides a measure of the bandpass-filter bandwidth, because one may uniquely identify small $r$ with a large bandwidth and $r \lesssim 1/4$ with a narrow bandwidth if one assumes that  $\omega_{h} \le \omega_{l}$ holds.

The trivial steady-state solution, i.e. $x=y=0$, is the only fixed point of model (\ref{model}). 
The local stability of the fixed point can be studied for general nonlinear functions by considering the Taylor-expansion of $f(x)$ around $x=0$.  Retaining the lowest-order term of the Taylor-expansion and multiplying by the gain factor $\gamma$ yields the effective slope, one of the relevant bifurcation parameters. Let us therefore introduce
\be
\label{def_b}
b = \gamma f'(0). 
\ee
The value of the effective slope $b$ where the fixed point first becomes linearly unstable is called the critical value, denoted as $b_{C}$. It determines the critical gain through $\gamma_C=f'(0)^{-1} b_{C}$.

For the discussion of linear stability and for the center-manifold analysis it is useful to rewrite (\ref{model}) as a functional differential equation and to explicitly separate the linear and nonlinear parts. Model (\ref{model}) may be written as
\be
\dot{u}(t)=L_{\mu}(u_t) + F(u_t,\mu) \qquad (t \ge 0),
\label{rfdemodel}
\ee 
where we use the following notation.
Let $u(t) \in \Reals^2$ be the vector $u=(x,y)^T$, and let $C=C\left([-\tau,0],\Reals^2 \right)$ be the Banach space of continuous functions mapping the interval $[-\tau,0]$ ($\tau>0$) into $\Reals^2$. For $\varphi \in C$ the norm is defined as $\| \varphi \|=\sup_{-\tau \le \theta \le 0}|\varphi(\theta))|$, where $|\cdot|$ is the norm in $\Reals^2$. Furthermore, for each fixed $t$, $u_t$ designates the function in $C$ given by $u_t(\theta)=u(t+\theta), \, \theta \in [-\tau,0]$. 
The parameter $\mu$ is defined through $\mu=b-b_{C}$, i.e. it is the bifurcation parameter shifted such that $\mu=0$ at the critical value.
$L_{\mu}: C \rightarrow \Reals^2 $ is a bounded linear operator and $F$ is a $C^k$ function containing only nonlinear terms, i.e. $F(0,\mu)=0$ and the first derivative $D_1F(0,\mu)=0$.
The linear operator $L_{\mu}$ is given by 
\be
\label{Lop}
L_{\mu}(\varphi) = \int_{-\tau}^0 d\eta(\theta,\mu) \varphi(\theta) := L_0(\varphi) + \mu L_1(\varphi),
\ee
where $\eta$ is a $n \times n$ matrix-valued function of bounded variation on $[-\tau,0]$, $\varphi \in C \left([-\tau,0],\Reals^2\right)$, and 
\be
L_0 (\varphi) = \left(\begin{matrix}
           -1 & 1 \\
           -r  &  0 \\
\end{matrix}\right)    \varphi(0)
+
\left(\begin{matrix}
          b_{C} & 0 \\
           0  &  0 \\
\end{matrix}\right)    \varphi(-\tau)
,
\qquad
L_1 (\varphi) =
\left(\begin{matrix}
           1  & 0 \\
           0  &  0 \\
\end{matrix}\right)    \varphi(-\tau).
\label{L0_L1}
\ee
The function $F$ is given by
\be
\label{F}
F(\varphi,\mu) 	= \left(\begin{matrix} 1 \\ 0 \end{matrix}\right) \left\{  \frac{1}{2}\, b_2(\mu)\, \left[ \varphi^{(1)}(-\tau) \right]^2 + \frac{1}{3!}\,b_3(\mu)\,\left[\varphi^{(1)}(-\tau)\right]^3 + h.o.t.  \right\} ,
\ee
where $\varphi^{(1)}$ denotes the first component of the vector $\varphi=(\varphi^{(1)},\varphi^{(2)})^T$ and the coefficients $b_2(\mu)$ and $b_3(\mu)$ are related to the Taylor-expansion of the nonlinearity $f(x)$ through the relations
\begin{equation}
\label{def_b2_b3}
\begin{split}
b_2(\mu) &= \gamma f''(0) = (b_{C} + \mu) \frac{f''(0)}{f'(0)}, \\
b_3(\mu) &= \gamma f'''(0) = (b_{C} + \mu) \frac{f'''(0)}{f'(0)}.
\end{split}
\end{equation}

We will use the abstract description of time-delay systems with band-limited feedback given by Eq. (\ref{rfdemodel})-(\ref{def_b2_b3}) to determine the stability of the fixed point $u=0$ and to derive the Hopf normal-form. However, we will refer to the more concrete description provided by model (\ref{model}) when we state results. This will make it easier to apply our findings to experiments.

\section{Global Stability of the fixed point for small gain}
\label{sec:globstab}

In the discussion of stability properties of solutions to dynamical systems the notions of global and local stability have to be distinguished. Global stability implies that the fixed point is the only attractor of the system, whereas local stability means that, after small perturbations away from the fixed point, the system will return to it's original state.
In this section, a sufficient condition for global stability of the steady-state solution of delay systems with band-limited feedback is provided. 

Consider the case $\gamma=0$ (no feedback) reducing (\ref{model}) to the second-order equation of a damped harmonic oscillator. In this case, the fixed point is globally stable. For small amounts of feedback one would expect that the dissipation in the system will dominate and the fixed point will as well be the global attractor. Under some assumptions on $f$ we can show that this is true (see Appendix \ref{proof-globstab} for the proof). 

\begin{prop} 
\label{stable_fp}
 Consider (\ref{model}) with $f(x)$ being continuous, Lipschitz in $x$, and in particular assume that there exists a constant $k_f$ such that $k_f |x| > |f(x)|$ for all $x \ne 0$. If $|\gamma|<k_f^{-1}$, the trivial solution of (\ref{model}) is globally uniformly asymptotically stable.
\end{prop}

The assumption that $f$ is Lipschitz is not very restrictive because
most physically implementable nonlinearities are differentiable (have a bounded derivative) and are therefore Lipschitz.  For these systems it is always possible to find a small but finite gain $|\gamma|$ for which the condition  $|x| > |\gamma| |f(x)|$ ($ \forall \, x \ne 0$) of Proposition \ref{stable_fp} is satisfied. However, in many cases the steady-state solution will be globally stable for a larger range of $|\gamma|$ because Proposition~\ref{stable_fp} only provides a sufficient condition. Furthermore, the bound was derived so that it holds for arbitrary delays $\tau$ and therefore may not be optimal. 
Nevertheless, the bound is both sufficient and necessary for nonlinearities $f$ for which $k_f$ is minimal, i.e. $k_f=|f'(0)|$, and thus global stability extends to $|b|=|\gamma f'(0)|<1$. The reason Proposition~\ref{stable_fp} provides as well the necessary condition for these nonlinearities is that, for $|b|>1$, there exist values of $\tau$ for which the steady state is locally unstable as will be shown in the next section.

\section{Local Stability of the Fixed Point}
\label{sec:locstab}

To gain a better understanding of how different dynamics arise as system parameters are varied, the local stability of the fixed point is discussed in this section using linear stability analysis. The main idea is to investigate how small perturbations to the trivial solution evolve for a given set of parameter values, which is equivalent to studying the corresponding characteristic equation. The goal is to construct a bifurcation plot showing the location of bifurcations as parameters are varied.
When a system has two or more parameters, as is the case here, it is common to draw two-parameter bifurcation plots. In this context, codimension-one bifurcations are located on one-dimensional curves and codimension-two bifurcations occur at isolated points in the bifurcation plot.

\subsection{Characteristic Equation}

For (\ref{rfdemodel}), the local stability of the fixed point is determined by the eigenvalues (or characteristic values) of the linearized equation, $\dot{u}(t)=L_{\mu}(u_t)$. The eigenvalues $\lambda$ satisfy the characteristic equation \cite{Hale2,Hale}
\be
\det \Delta(\lambda,0)=0, \qquad \Delta(\lambda,\mu)=\lambda \id - L_{\mu}\left(e^\lambda \id \right).
\label{characteristic-abstract}
\ee
and coincide with those values of $\lambda$ for which there is a nonzero vector $c_{\lambda}$ such that there is a solution of $\dot{u}(t)=L_{\mu}(u_t)$ given by $u_t=c_{\lambda} \exp(\lambda t)$. With
$L_{\mu}$ defined by (\ref{Lop}), the characteristic equation (\ref{characteristic-abstract}) is given by
\be
 \lambda^2  +\lambda + r - b \lambda e^{-\lambda \tau} =0.
\label{characteristic1}
\ee
To determine the local stability of the fixed point, we first consider solutions of (\ref{characteristic1}), which is transcendental and has an infinite number of roots for every set of fixed parameter values. The fixed point is locally stable if all roots (eigenvalues) have a negative real part. Thus, to determine  parameter values for which the fixed point becomes unstable, we set $\Real (\lambda)=0$ and $\Imag (\lambda)=i \Omega$. Separating real and imaginary part yields
\be\label{characteristic2}
\begin{split}
0&=-\Omega^2+r-b \Omega \sin( \Omega \tau) \\
0&= 1 - b \cos(\Omega \tau). \\
\end{split}
\ee
Note that this set of equations is unchanged for $\Omega \rightarrow -\Omega$ and that there is no solution for $\Omega=0$ if $r>0$. This implies that a pair (or pairs) of complex-conjugate eigenvalues will cross the imaginary axis as the bifurcation parameter is varied. Thus, it is sufficient to consider positive $\Omega$ only and, generically, the trivial solution becomes unstable through a Hopf bifurcation, which is a codimension-one bifurcation.  

One way to visualize the solution of (\ref{characteristic2}) is to seek parameterized curves in the plane of two bifurcation parameters, which we choose as the delay $\tau$ and the effective slope $b$. These curves separate regions in parameter space with different numbers of eigenvalues in the right complex halfplane. The relevant region where the fixed point is stable (no eigenvalues in the right complex halfplane) is the one that includes $\gamma=b=0$.  The parameterization of the curves is most conveniently achieved through $s=\Omega \tau$ yielding
\begin{xalignat}{1}
\tau_C^n(s)   &=\frac{s}{2 r} \left(\tan(s) + \sqrt{ \tan^2(s)+ 4 r } \right)
\label{tauc} , \\
b_{C}^n(s) &= \frac{1}{\cos(s)} \label{bc}.
\end{xalignat}
Here, the label $n$ denotes the different solution branches and the value of $n$ is defined in terms of the curve parametrization as
\begin{xalignat}{2}
0 &< s < \frac{\pi}{2} &\Leftrightarrow & \qquad n=0 ,\label{ntox_1} \\
\frac{(2 n -1)\pi}{2} &< s < \frac{(2n+1) \pi}{2}  &\Leftrightarrow &\qquad n=1,2,3,\ldots  .
 \label{ntox_2}
\end{xalignat}
Note that $b_{C}^n$ is positive for even $n$ and negative for odd $n$.
Also, note that $\tau_C^0$ approaches zero as $s \rightarrow 0$. Furthermore, $\tau_C^n \rightarrow 0$ as $s$ approaches $ (2 n-1) \pi / 2$ from above but $\tau_C^n$ tends to infinity as $s$ approaches $ (2 n+1) \pi / 2$ from below.

\begin{figure}
 \centerline{\includegraphics[width=\columnwidth]{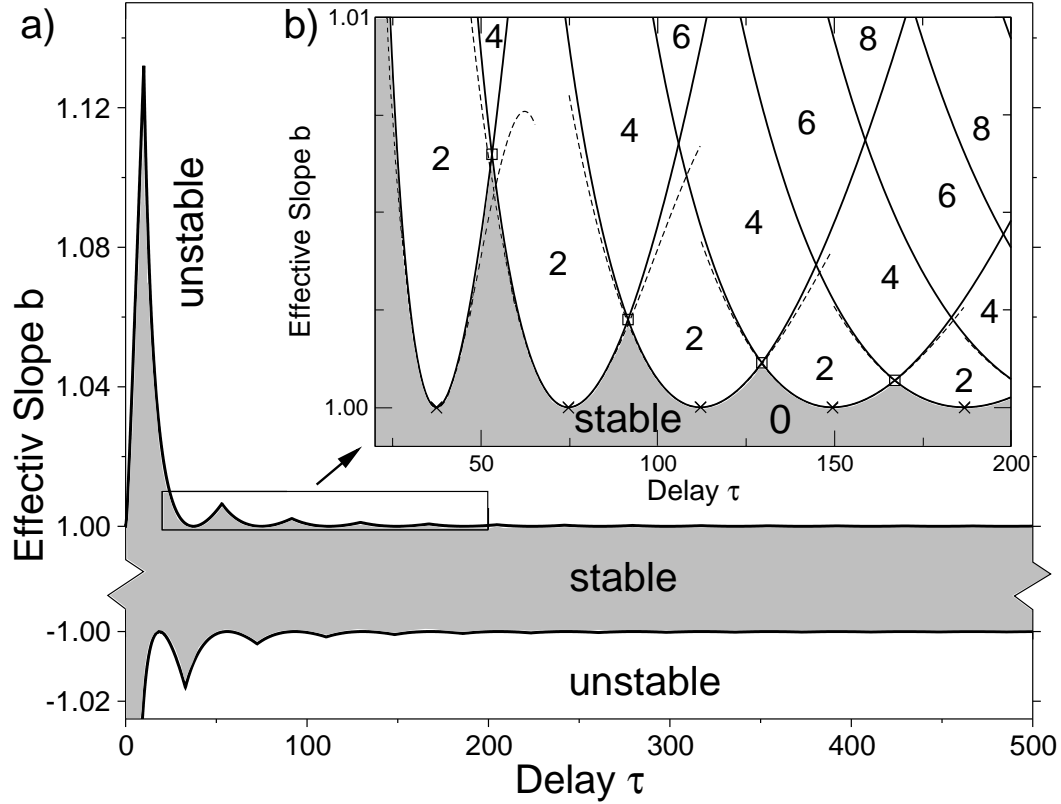}}
   \caption{Local stability of the fixed point, indicated by the shaded region, as a function of the parameters $\tau$ and $b$. a) A large part of the stability region ($-1<b<1$) has been contracted to make details of the boundary visible.
b) The inset shows a portion of the stability boundary. The number of eigenvalues in the right complex halfplane is given.
 The square symbols mark points on the stability boundary where two pairs of complex conjugate eigenvalues lie on the imaginary axis, whereas crosses denote the extrema of the Hopf-curves, i.e. $b=1$. The dashed lines depict the approximation given by (\ref{bcapprox}).}
\label{fig:stabbdry}
\end{figure} 

In Fig.~\ref{fig:stabbdry}, the stability boundary of the fixed point is shown in $\tau$-$b$-parameter space. The stability boundary is obtained by combining those pieces of the parameterized curves $(\tau_C^n(s),b_{C}^n(s))$, given by (\ref{tauc}) and (\ref{bc}), for which, upon crossing, the number of eigenvalues in the right complex halfplane changes from zero to two. From Fig.~\ref{fig:stabbdry}a it is seen that the fixed point is always locally stable if $-1< b <1$, that it may be stable for a considerably larger range of $b$ if $\tau$ is small, and that $|b_{C}| \approx 1$  if $\tau$ is large.

Figure~\ref{fig:stabbdry}b provides a zoomed view of part of the parameter space where the parameterized curves $(\tau_C^n(s),b_{C}^n(s))$ are shown as solid lines (starting from the left the curve-index $n$ is $n=2,4,6,8,\ldots$). We will refer to these curves as Hopf-curves. 
  As one of the Hopf-curves in Fig.~\ref{fig:stabbdry}b is crossed from below, a pair of complex conjugate eigenvalues moves into the right complex halfplane, which is seen by considering how the eigenvalues $\lambda$ change with respect to $b$ for fixed $\tau$. 
Using (\ref{characteristic1}) we calculate the real part of the derivative of $\lambda$ at the critical point,
\be
\label{dlambda_answer}
 \Real \left.\frac{d \lambda}{d b} \right|_{critical}= b^n_{C} \;
\frac{\tau_C^n\, s \left( s +\cos(s) \sin(s) \right)   + 2 s^2  \cos^2(s) }
{| 2 s + \tau_C^n \tan(s) + \tau_C^n \, s \cos(s)^{-1} e^{-i s} |^2 }.
\ee
Because $s=\Omega \tau>0$ implies that $s +\cos(s) \sin(s)>0$, it follows that the right hand side of (\ref{dlambda_answer}) is always nonzero and that it's sign is determined by the sign of $b^n_{C}$. Therefore, for branches with positive $b^n_{C}$, the eigenvalues move into the right complex halfplane as $b$ is increased, whereas  they move into the right complex halfplane for branches with negative $b_{C}^n$ as $b$ becomes more negative.

The above discussion confirms that the steady state will, generically, become unstable through a Hopf bifurcation in systems with band-limited feedback. Let us make this statement more precise. The two conditions for a Hopf bifurcation are that there is a pair of simple characteristic roots crossing the imaginary axis transversally at $\mu=0$ ($b=b_C$) and that the characteristic equation $\Delta(\lambda,0)$ has no other roots with zero real parts. 
 The roots (eigenvalues) cross transversally because the right hand side of (\ref{dlambda_answer}) is always nonzero. Furthermore, there is exactly one pair of roots on the imaginary axis for all parameter combinations of $b$ and $\tau$ that fall on one of the Hopf-curves. The exception are points where two Hopf-curves intersect, because in that case two pairs of eigenvalues cross into the right halfplane and a codimension-two bifurcation (double-Hopf bifurcation) occurs. 
Let us denote as $\tau^+$ ($\tau^-$) the set of delays $\tau$ at which the fixed point becomes unstable due to such a codimension-two bifurcation  as the effective slope $b$ is increased (decreased).
Because the sets $\tau^+$ and $\tau^-$ are countably infinite, i.e. have Lebesgue measure zero on $\Reals^+$, the fixed point loses stability through a Hopf-bifurcation for essentially all delays $\tau$.  In other words, careful tuning of the gain ($ \propto b$)  and the delay ($\tau$) is necessary to bring an experiment close to a codimension-two bifurcation point. It is in this sense that the Hopf bifurcation is generic for time-delay systems with band-limited feedback.

\subsection{Approximate analytic solution of the characteristic equation}

Although (\ref{tauc}) and (\ref{bc}) allow us to determine numerically the critical parameters to a desired precision, it is  useful to have an explicit expressions for the critical value of the bifurcation parameter. For many experiments the feedback gain is most easily varied, which means that the transcendental equations (\ref{tauc}) and (\ref{bc}) should be inverted to give $b_{C}^n$ as a function of the remaining two parameters $\tau$ and $r$.

An exact solution is not possible in general. Therefore, we solve approximately around the extrema of the stability boundary, i.e. around points along the Hopf-curves where $s=n \pi$ (crosses in Fig.\ref{fig:stabbdry}):
\begin{xalignat}{3}
  \tau_C^n(s=n\pi) &= \frac{n \pi}{\sqrt{r}},& 
  b_{C}^n(s=n\pi)&=\pm 1, & 
  \Omega^n_C (s=n\pi)&= \sqrt{r}.
\label{extrema}
\end{xalignat}
We expand the trigonometric functions and the square root in (\ref{tauc}) and (\ref{bc}) around the extrema and solve for the critical effective slope  
\be
b_{C}^n(\tau,r) = \pm \left(1 + \frac{2 r^2 \left(\tau - \frac{n \pi}{\sqrt{r}}\right)^2 }{(n \pi+ 2 \sqrt{r})^2} - \frac{2 (n \pi r^{5/2} + 4 r^3) \left(\tau - \frac{n \pi}{\sqrt{r}}\right)^3}{(n \pi + 2 \sqrt{r})^4} \right) + h.o.t.
 \label{bcapprox}
\ee
The approximation given by (\ref{bcapprox}) is shown in  Fig.~\ref{fig:stabbdry} using dashed lines.

\subsection{The frequency at the onset of instability}

\begin{figure}
   \centerline{\includegraphics[width=\columnwidth]{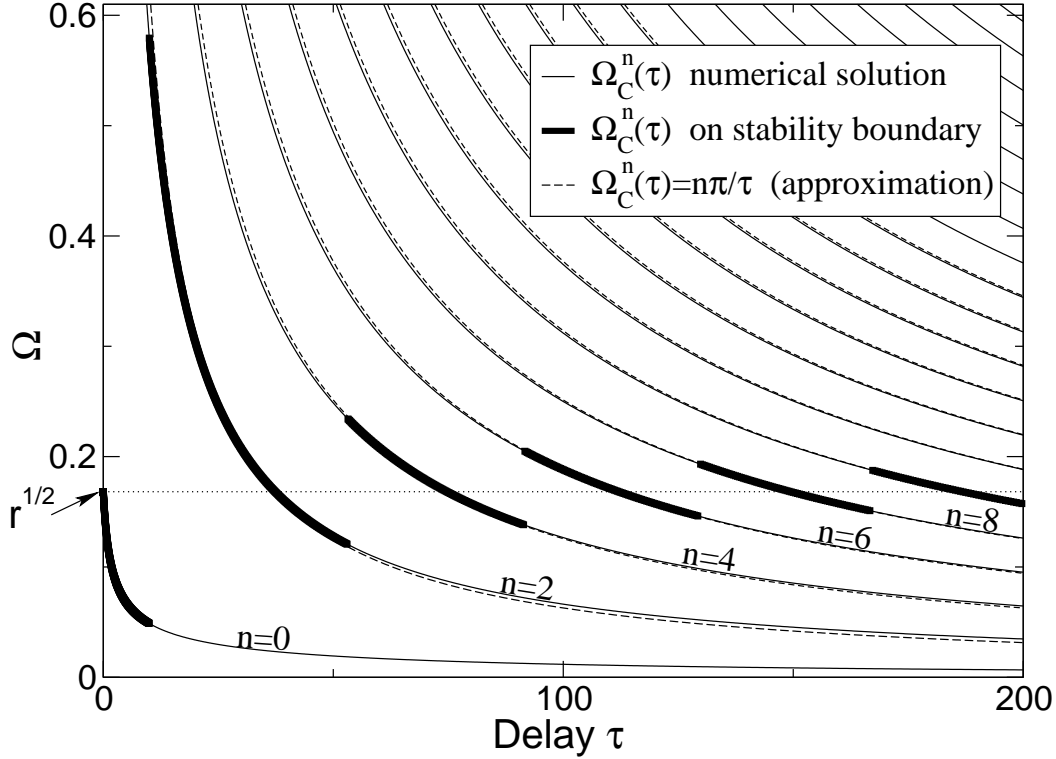}}
   \caption{Imaginary part of the eigenvalue versus the delay. $\Omega^n_C(\tau)$ is shown for positive $b$ ($n$ is even) and $r=0.028$ . The part of each branch $n$ that belongs to the stability boundary is shown as bold straight line. The dashed lines are the approximate value of the frequency.  }
\label{fig:omega}
\end{figure} 

So far we have discussed the values of the feedback gain for which the steady state is stable. That is, we have shown how to calculate the critical gain $\gamma_C$ for known values of the delay  ($\tau$), the first derivative of the nonlinearity ($f'(0)$), and the frequency of maximal transmission ($\sqrt{r}$). 
The value of $\gamma_C$ can be compared with experiments. Another useful way to compare experiments and  theory is to study the frequency of oscillations at the onset of instability.

Consider the case where the Hopf bifurcation is supercritical. Then, as the stability boundary is crossed, the fixed point becomes unstable, a stable limit cycle is born, and the system starts to oscillate. Experimentally, the frequency of this oscillation is the quantity that is most readily measured.  The frequency at onset $f^n_C$ is determined by $\Omega^n_C$ through $f^n_C=\Omega^n_C (2 \pi)^{-1}(\omega_h+\omega_l)$. Therefore, it is useful to plot the eigenvalue $\Omega^n_C$ versus the delay $\tau$ as is done in Fig.~\ref{fig:omega}.
 
In the vicinity of the extrema (Eq.~(\ref{extrema}), crosses in Fig.~\ref{fig:stabbdry}), the frequency scales roughly as $\Omega^n_C \sim n \pi/\tau$ ($f_C^n \sim n/ (2 T)$) for $n>0$.
This approximate behavior, shown as dashed curves in Figure \ref{fig:omega}, can be understood by considering whether a wave circulating in the feedback loop will reinforce itself. For the case of positive feedback ($b>0$) a periodic perturbation will reinforce itself, if the feedback delay is a multiple of the wave's period, i.e. $f \sim n / (2 T)$ with $n$ an even integer.  On the other hand, a sinusoidal perturbation is amplified by negative feedback ($b<0$), if it is shifted by half it's period after one round-trip. Thus, for $b_C<0$ the frequency is expected to scale as $f \sim n / (2 T)$ with $n$ an odd integer. This reasoning is consistent with the fact that for $n$ even (odd) the critical effective slope $b_C^n$ is positive (negative).

From Fig.~\ref{fig:omega}, it is also seen
that different oscillation ``modes'' will be observed as $\tau$ is increased. That is, the frequency of the observed oscillations will jump  from $\Omega \sim n \pi/ \tau$ to $\Omega \sim (n+2) \pi / \tau$ as $\tau$ is varied across one of the double-Hopf points ($\tau \in \tau^{\pm}$; squares in Fig.~\ref{fig:stabbdry} inset).
These jumps are explained by the fact that the gain in the feedback loop is not perfectly flat over the pass-band.  As $\gamma$ is increased from a low level, one particular frequency will first reach the threshold where the gain in the loop balances the losses. In a system with only low-pass feedback, the gain is highest at low frequencies, so the oscillation-mode with the lowest frequency is always the one that destabilizes the steady state, independent of the delay. On the other hand, the high-pass filter introduces a bias toward high frequencies. Because the frequency scales roughly as $\Omega \sim \tau^{-1}$ for each mode $n$, the damping effect of the high pass filter on a particular mode becomes more pronounced with increasing delay time $\tau$. Therefore, there exists a delay $\tau$ for which a higher order mode, one that has a higher frequency for a given delay, will reach threshold first.
Thus, examining the frequency of the oscillations at the onset of instability as a function of the delay allows one to distinguish time-delay systems with band-limited feedback from systems with low-pass feedback \cite{laspaper}.

 For completeness, note that $\Omega^n_C$ can be determined exactly if $\tau$ is considered as bifurcation parameter and $b$ is kept constant, and is given by
\be
(\Omega_C^n)^2 = 
  \begin{cases}
    \frac{1}{2}\left\{[(b)^2+2r-1] + \sqrt{ [(b)^2+2r-1]^2 - 4 r^2 } \right\} & \text{ if } \tau_C^n \le \frac{n \pi}{\sqrt{r}} , \\
  \frac{1}{2}\left\{ [(b)^2+2r-1] - \sqrt{[(b)^2+2r-1]^2  - 4 r^2 } \right\}& \text{ if } \tau_C^n > \frac{n \pi}{\sqrt{r}} , 
  \end{cases}
 \label{Omegaofb1}
\ee
as is discussed in detail in \cite{Kuang}.

\section{Hopf Normal-Form}
\label{sec:normal_form}

In the previous sections, we have shown that a time-delay system with band-limited feedback undergoes Hopf bifurcations as the system parameters are varied. However, we have not yet determined whether the
Hopf bifurcations are  subcritical or supercritical. The expected dynamical behavior close to threshold is very distinct for the two bifurcation types. 
 For a supercritical Hopf bifurcation, one expects small amplitude sinusoidal oscillations past the critical gain, whereas for a subcritical bifurcation, one expects bistability and hysteresis close to the bifurcation point.
This section is devoted to the determination of the type of the Hopf bifurcation using normal form theory. 

Normal forms have been studied extensively for finite-dimensional ordinary differential equations (ODEs) \cite{guckenheimer}.
Recently, Faria and  Magalh\~{a}es  developed normal-form theory for the case of DDEs  \cite{Faria2,Faria1}. 
As for ODEs, one tries to simplify the dynamical systems under study using two approaches: reduce the dimension of the system and remove as much of the nonlinearity as possible. The first goal is achieved by considering the restriction of the flow to the center manifold. 
This restriction is justified because the qualitative dynamical behavior of solutions of the infinite-dimensional DDE close to the fixed point can be described by the finite-dimensional ODE associated with the flow on the center manifold.
The second goal is achieved by transforming the nonlinear differential equation into an equation with a simpler algebraic form, called a normal form, by changes of variables that eliminate all irrelevant terms from the equation, but retain the qualitative properties of the flow.

The normal form theory developed in Refs.~\cite{Faria2,Faria1} uses the formal adjoint theory for linear DDEs \cite{Hale2} to set up an appropriate coordinate system near an equilibrium point. It then proceeds by performing a sequence of transformations of variables such that, at each step $j$, the change of variables effects simultaneously the projection of the original DDE on the center manifold and removes the non-resonant terms of order $j$ from the ODE on it.

For a Hopf bifurcation, the normal form on the center manifold is given in polar coordinates $\rho,\xi$ by
\begin{equation}
\begin{split}
\dot \rho &= \mu \, K_1 \rho + K_2 \rho^3 + \O(\mu^2 \rho + |(\mu,\rho)|^4) \\
\dot \xi  &= - \Omega_C^n + \O(|(\mu,\rho)|).
\end{split}
\label{normal_form}
\end{equation}
Here, $\mu=b-b_C^n$ is the bifurcation parameter, $\Omega_C^n$ is the imaginary part of the eigenvalue, and $K_1$ and $K_2$ are real coefficients. 
The qualitative behavior of the asymptotic solutions of (\ref{rfdemodel}) (or equivalently of (\ref{model})) is the same as the behavior of solutions of (\ref{normal_form}), which, in turn, only depends on the signs of the two coefficients $K_1$ and $K_2$. Of special interest is to distinguish between the cases $K_2<0$ indicating a supercritical bifurcation, and $K_2>0$ indicating a subcritical bifurcation. The main object of this section is to express $K_1$ and $K_2$ in terms of the relevant parameters ($b$, $\tau$, $r$, and the Taylor-expansion coefficients of the nonlinearity $F$) and to determine the sign of $K_1$ and $K_2$.

\subsection{Derivation of the normal form coefficients}

For scalar DDEs (delay systems with low-pass feedback) an explicit formula for $K_1$ and $K_2$ is given in Ref. \cite{Faria2} and reproduced in \cite{Hale} (8.3). Based on this work, Giannakopoulos and Zapp \cite{Zapp} discuss in detail the different possible scenarios of Hopf bifurcations in scalar DDEs. As far as we are aware, no analogous work exists for two-dimensional DDEs.
Nevertheless, we find that the derivation of the formula for  $K_1$ and $K_2$ for the two-dimensional case carries through entirely analogous to the derivation given in Refs.~\cite{Hale,Faria2}. We will therefore restrict ourselves to the definition of the necessary quantities and to the statement of the end result. We adopt a notation that follows closely that of Ref.~\cite{Hale}.

 We take $\mu$, or equivalently the effective slope $b$, as our bifurcation parameter. The other parameters, e.g. $\tau$ and $r$, are assumed to be fixed.
 For a given set of parameters $\tau$ and $r$, there are an infinite number of critical effective slopes $b_C^n$, distinguished here by the solution-branch label $n$ (see (\ref{ntox_1}) and (\ref{ntox_2})). 
For each $b_C^n$, there exists a center manifold at $u=0$ and the dynamics of (\ref{rfdemodel}), restricted to this manifold, is described by (\ref{normal_form}). That is, the derivation of the normal form is the same for all $n$.
Furthermore, all expansion coefficients are evaluated at the bifurcation point, $\mu=0$. Consequently, to simplify notation,
we will suppress explicit reference to $n$ and $\mu$ for the rest of the section and write $b:=b_{C}^n$, $\tau:=\tau_C^n$, $\Omega:=\Omega_C^n$, $b_2:=b_2(\mu=0)$, and $b_3:=b_3(\mu=0)$. 

The first step in applying center-manifold theory to our problem is to introduce suitable local coordinates. For the Hopf-bifurcation it is convenient to work with complex variables, i.e to work in $C= C \left([-\tau,0],\Com^2\right)$.
 Considering the linearized flow
\be
\dot{u}(t)=L_{0}(u_t) \qquad (t \ge 0),
\label{rfdemodel_lin}
\ee 
  one can associate with each fixed set of eigenvalues $\Lambda=\{ \lambda_1,\ldots,\lambda_s \}$ a generalized eigenspace $P$ with base $\Phi$ consisting of the appropriate eigenfunctions. 
In our case $\Lambda=\{ i \Omega, - i \Omega \}$, and, denoting the complex conjugate using an overline, the base $\Phi$ can be chosen as  $\Phi = ( \phi_1, \phi_2 )$ with $\phi_2=\overline{\phi_1}$, where 
\be
\label{base}
 \phi_1(\theta) =
\left(\begin{matrix}
         1 \\
        i r / \Omega  \\
\end{matrix}\right)  e^{i \Omega \, \theta} \quad \text{with} \quad \phi_1 \in C \left([-\tau,0],\Com^2\right), \quad -\tau \le \theta \le 0. 
\ee
Using this notation, it follows that $P=\text{span} \, \Phi$. To compute projections onto this generalized eigenspace one needs to consider the adjoint equation (see \cite{Hale2} for a definition).
Let $\Com^{2*}$ be the space of row 2-vectors, $C^*:=\left([0,\tau],\Com^{2*}\right)$, and let $P^*$ denote the dual of $P$. The base of $P^*$ is $\Psi = col ( \psi_1, \psi_2 )$ with $\psi_2=\overline{\psi_1}$ and
\be
\label{dualbase}
\psi_1(\zeta)=  N \left( \begin{matrix} 1, &  -i/\Omega \end{matrix} \right) e^{-i \Omega \zeta} \quad \text{with} \quad \psi_1 \in C^* \left([0,\tau],\Com^{*2}\right), \quad 0 \le \zeta \le \tau. 
\ee  
The normalization factor $N$ is chosen to be
\be
\label{normfactor}
N=\left( 1 + b \tau e^{-i \Omega \tau} + \frac{r}{\Omega^2} \right)^{-1}.
\ee
The formal duality is the bilinear form $\left( \cdot| \cdot \right)$ from $C^* \times C$ to the scalar field \cite{Hale2,Hale}
\be
\left( \alpha| \varphi \right) = \alpha(0) \varphi(0) - \int^0_{\theta=-\tau}  \int_{\xi=0}^{\theta}  \alpha(\xi-\theta)\; d\eta(\theta)\; \varphi(\xi) \;d\xi \quad \alpha \in C^*,\; \varphi \in C,
\ee
which is simplified by using the definition of $\eta$ provided by $(\ref{Lop})$
\be
\label{scalarprod}
\left( \alpha| \varphi \right) = \alpha(0) \varphi(0) + \int^0_{\theta=-\tau} \alpha(\xi+\theta) \;
\left(\begin{matrix}
          b & 0 \\
           0  &  0 \\
\end{matrix}\right) \; \varphi(\xi) \;d\xi.
\ee
The normalization factor $N$ (\ref{normfactor}) is chosen such that $\left( \Psi| \Phi \right) = \id$, where $\id$ is the identity matrix. The purpose of above definitions is that it allows one to decompose the phase space $C$ into the generalized eigenspace of (\ref{rfdemodel_lin}) associated with $\Lambda$, i.e. $P$, and $Q=\{ \varphi \in C : \left(\Psi | \varphi \right) = 0 \}$. In short, the phase space $C$ is decomposed by $\Lambda$ as $C= P \oplus Q$. 

In Ref.~\cite{Faria1}, it is shown that the phase space associated with the full nonlinear problem (\ref{rfdemodel}) may similarly be decomposed and a projection from that phase space to $P$ is given in terms of the bases $\Phi$ and $\Psi$. Furthermore, the center manifold theorem assures that there is a 2-dimensional invariant manifold of (\ref{rfdemodel}) tangent to the center space $P$ of (\ref{rfdemodel_lin}) at zero. It is therefore possible to use this projection and appropriate coordinate transformations to obtain an ODE on the center manifold in normal form, i.e. (\ref{normal_form}). We refer the reader to Ref.~\cite{Hale}, Sec. 8.3 for details.

Since the Hopf bifurcation is  determined generically up to third order (see (\ref{normal_form})) we need the Taylor-expansion of the nonlinearity $F$ up to third order. Write the Taylor expansion of $F$ at the critical point as
\be
F(u_t,\mu)=F(u_t,0) = \frac{1}{2} F_2(u_t,0) + \frac{1}{3!} F_3(u_t,0) + h.o.t.
\ee
Let $H_2: C \times C \rightarrow \Com^2$ be the bilinear symmetric form such that $F(u,\mu)=F(u,0)=H_2(u,u)$,
\be
\label{H2}
H_2(u,v) = b_2 \,
\left(\begin{smallmatrix}
          1 \\
          0 \\
\end{smallmatrix}\right)
u^T(-\tau)
\left(\begin{smallmatrix}
          1 & 0 \\
          0 & 0 \\
\end{smallmatrix}\right)
v(-\tau),
\ee
where  the definition of the nonlinearity $F$ (Eq.~(\ref{F})) is used. 

Denote the complex coordinates on the two-dimensional center manifold by $z$, and the components of $z$ by $z_1$ and $z_2$. These coordinates are related to the polar coordinates of the normal form (\ref{normal_form}) through $z_1=\rho \cos \xi - i \rho \sin \xi$ and $z_2=\overline{z_1}$.
In the computation of the coefficients $K_1$ and $K_2$, the expansion of $F(\Phi z,0)$ plays an important role. For the second-order terms write
\be
F_2(\Phi z,0)= A_{(2,0,0)} z_1^2 + A_{(1,1,0)} z_1 z_2 + A_{(0,2,0)} z_2^2. 
\ee
 The coefficients are obtained by using the definition of $\Phi$, (\ref{base}), and $F$, (\ref{F}):
\be
A_{(2,0,0)} = H_2(\phi_1,\phi_1)=\left(\begin{smallmatrix}
          1 \\
          0 \\
\end{smallmatrix}\right)
b_2 e^{-2 i \Omega \tau} = \overline{A_{(0,2,0)}}
\label{A200},
\ee
\be
A_{(1,1,0)} = 2 H_2(\phi_1,\phi_2)=\left(\begin{smallmatrix}
          1 \\
          0 \\
\end{smallmatrix}\right)
2 b_2 \qquad \qquad \qquad
\label{A110}.
\ee

Write the third-order terms as
\be
F_3(\Phi z, \mu)= \sum_{q_1+q_2+l=3} A_{(q_1,q_2,l)} z_1^{q_1} z_2^{q_2} \mu^l. 
\ee
 For the Hopf bifurcation, the only relevant third-order coefficient is
\be
A_{(2,1,0)} = \left(\begin{smallmatrix}
          1 \\
          0 \\
\end{smallmatrix}\right)
3 b_3 e^{-i \Omega \tau}
\label{A210}.
\ee

With this result, we are finally able to state the main result of this section.
Our calculation of $K_1$ yields
\be
K_1 = \Real\left( \psi_1(0) L_1(\phi_1) \right) ,
\ee
and for $K_2$ we obtain
\bea
K_2 &=& \frac{1}{3!} \Real\left( \psi_1(0) A_{(2,1,0)} \right)
       + \frac{1}{2}  \Real\left\{ \psi_1(0) H_2\left( \phi_1, \left[ - L_{0}\left(\id \right) \right]^{-1}  A_{(1,1,0)} \right)   \right\}     
\nonumber \\
  &+&
     \frac{1}{2} Re\left\{ \psi_1(0) H_2\left( \overline{\phi_1}, \left[  2 i \Omega \id - L_{0}\left(e^{2 i \Omega \theta} \id \right) \right]^{-1}  A_{(2,0,0)} e^{2 i \Omega \theta} \right)   \right\}.     
\label{K2_general}
\eea
The above expressions are quite general \cite{generality} and are an obvious extension of the formulas for the Hopf normal-form coefficients for scalar DDEs given in \cite{Hale}, pp. 154, Eq. (8.77). 

Next, we express $K_1$ and $K_2$ explicitly in terms of the parameters that are specific to the problem at hand. 
First, note that the following identity holds
\be
K_1=  \Real \left.\frac{d \lambda}{d b} \right|_{critical}.
\ee
Thus, the coefficient $K_1$ is given by (\ref{dlambda_answer}), which implies that $K_1$ is positive for $b>0$ and negative for $b<0$ (see section Sec.~\ref{sec:locstab}).
Second, in (\ref{K2_general}) the use of the definition of $L_0$, (\ref{L0_L1}), and of $H_2$, (\ref{H2}), and the substitution of $A_{(2,1,0)}$, $A_{(2,0,0)}$, $A_{(1,1,0)}$, $\phi_1$, and $\psi_1$ (given, respectively, by 
(\ref{A200}), (\ref{A110}), (\ref{A210}), (\ref{base}), and (\ref{dualbase})) yields
\bea
K_2 &=& \frac{1}{2} \Real\left(\frac{b_3 \, \Omega^2}{ (b \tau \Omega^2 + (r + \Omega^2) e^{i \Omega \tau}) }  \right) + \nonumber \\
&& +
\Imag \left( \frac{(b_2)^2 \, \Omega^3}{ (b \tau \Omega^2 + (r + \Omega^2) e^{i \Omega \tau}) (2 i \Omega b + (4 \Omega^2 - 2 i \Omega - r) e^{2 i \Omega \tau} ) } \right) .
\label{K2}
\eea

\subsection{The sign of the normal-form coefficient $K_2$}

Equation (\ref{K2}) can be used directly to  determine numerically the sign of $K_2$ because the values of $\tau, r,\Omega$, and $b$ on each of the Hopf curves are known. However, since we are only concerned with the sign of $K_2$ it is possible to derive a somewhat simpler criterion for distinguishing  subcriticality from supercriticality. To that avail, define a function $C^n(\tau,r)$ through
\bea
C^n(\tau,r)&=&  \frac{- 2 \Omega^2  b^{-1} \; N^n }{ (b^2 \Omega^2 \tau + r+\Omega^2) \left|2 i \Omega b + (4 \Omega^2 - 2 i \Omega - r) e^{2 i \Omega \tau} \right|^2} \label{Cfunc}, \\
N^n&=&[ r + \Omega^2 ] [b^4 +2 b^3+b^2-4] + 2 \Omega^2 \tau b^2 [b^3-1] \nonumber \\
&& +\sgn(\tau-\frac{n \pi}{\sqrt{r}}) 3 \Omega \sqrt{b^2-1} [ (4-b^2)(r+\Omega^2) + 2 \Omega^2 \tau b^2 ] \label{Nfunc}.
\eea
Here, we restored the explicit reference to the label $n$ for clarity. Recall that  we use $b=b_C^n$ and $\Omega=\Omega_C^n$ in this section and that the numerical values of $b$ and $\Omega$ depend on the remaining parameters $\tau$ and $r$. It can be shown (see Appendix \ref{proof_main_theorem}) that  
\begin{prop}
\label{main_theorem}
For $r, \tau \in \Reals^+$, $r\le \frac{1}{4}$, and $n=0,1, 2,\ldots$ :
\begin{itemize}
\item If $f'(0) \, f'''(0) + f''(0)^2 \, C^n(\tau,r) > 0$, the Hopf bifurcation is subcritical ($K_2>0$).
\item If $f'(0) \, f'''(0) + f''(0)^2 \, C^n(\tau,r) < 0$, the Hopf bifurcation is supercritical ($K_2<0$). 
\end{itemize}
\end{prop}
Since the Taylor-series coefficients of $f$ are known, the difficulty in determining the sign of $K_2$ is shifted to finding the value of $C^n(\tau,r)$. Two simple cases that arise in this context are the following:
\begin{cor}
\label{cases_1_2}
\begin{itemize}
\item[]
\item[\normalfont (1)]
The case when there are no quadratic terms in the Taylor expansion of the nonlinearity ($f''(0)=0$ and $f'''(0) \ne 0$): If $f'(0) f'''(0) > 0$, the Hopf bifurcation is subcritical. If $f'(0) f'''(0) < 0$, the Hopf bifurcation is supercritical. In particular the type of bifurcation is independent of the sign of $\gamma$.
\item[\normalfont (2)]  The case when there are no cubic terms in the Taylor expansion of the nonlinearity ($f'''(0)=0$ and $f''(0) \ne 0$): The Hopf bifurcation is subcritical if $C^n(\tau,r)>0$ and supercritical if $C^n(\tau,r)<0$.
\end{itemize}
\end{cor}

In light of (2) in above Corollary, it is clearly useful to look for conditions on $\tau$ and $r$ for which the sign of $C^n(\tau,r)$ can be determined without explicit evaluation. To that avail we establish the following properties (see Appendix \ref{proof_of_C}): 
\begin{prop} 
\label{properties}
For $r, \tau \in \Reals^+$, $r\le \frac{1}{4}$, and $n=0,1, 2,\ldots$ :
\newline
$\bullet$ If $\tau > \frac{n \pi}{\sqrt{r}}$, $C^n(\tau,r)<0$. \newline
$\bullet$ If $\tau = \frac{n \pi}{\sqrt{r}}$ and $n$ even $($i.e. $b=1)$, $C^n(\tau,r)=0$. \newline
$\bullet$ If $\tau < \frac{n \pi}{\sqrt{r}}$ and $b \in (1,2]$, $C^n(\tau,r)>0$. \newline
$\bullet$ If $\tau \le \frac{n \pi}{\sqrt{r}}$ and $b \in [-2,-1]$, $C^n(\tau,r)<0$.
\end{prop}

Let us restrict our attention to the stability boundary of the fixed point, since this is most relevant for experiments. It is possible to show \cite{bbounds} that the stability boundary is bounded by $1 \le b < 2 $ for all delays $\tau$ ($\tau>0$) for positive effective slopes. For negative effective slopes the stability boundary is bounded by $-2 < b \le -1$ for $\tau > \tau_C^n(s)=\tau_C^1(4 \pi/3)$, where $\tau_C^n(s)$ is given by (\ref{tauc}).
Combining this with the results of Proposition~\ref{main_theorem} and Proposition~\ref{properties} the following simple cases arise in addition to those of Corollary \ref{cases_1_2}:
\begin{cor} At the stability boundary of the fixed point, the following holds for the Hopf bifurcation :
\label{cases_3_4_5}
\begin{itemize}
\item[\normalfont (3)]If $b<0$, $\tau > \tau_C^1(\frac{4 \pi}{3})$, and $f'(0) f'''(0) < 0$, the bifurcation is supercritical.
\item[\normalfont (4)] If $b > 0$, $\tau> \frac{n \pi}{\sqrt{r}}$, and $f'(0) f'''(0) < 0$, the bifurcation is supercritical.
\item[\normalfont (5)] If $b > 0$, $\tau< \frac{n \pi}{\sqrt{r}}$, and $f'(0) f'''(0) > 0$, the bifurcation is subcritical.
\end{itemize}
\end{cor}

In this section, we provide in Proposition \ref{main_theorem} a criterion that determines the type of Hopf bifurcation in general but requires the numerical evaluation of the function $C^n(\tau,r)$. The Hopf bifurcation type can be determined without the evaluation of $C^n(\tau,r)$ if certain assumptions about the parameters of model (\ref{model}) are satisfied. These simple cases are summarized in Corollary \ref{cases_1_2} and  Corollary \ref{cases_3_4_5}. To illustrate the findings of this section we discuss in the next section some specific examples of time-delay feedback systems with band-limited feedback.

\section{Examples}
\label{sec:examples}

\subsection{`Sine'-Nonlinearity}
%
%
\begin{figure}
    \centerline{\includegraphics[width=\columnwidth]{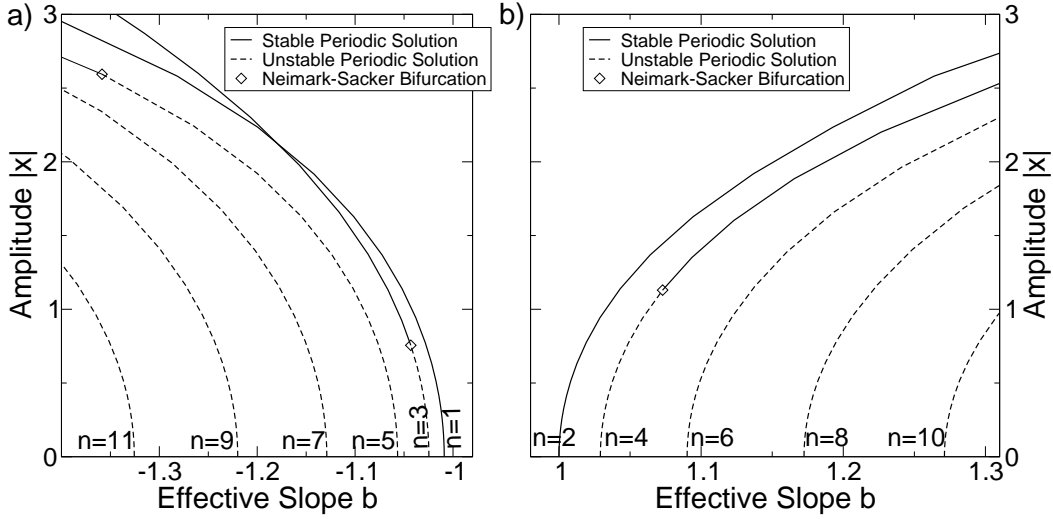}}
   \caption{The amplitude of the limit-cycle versus the effective slope $b$ is shown for $\tau=2 \pi/\sqrt{r}$ and $r=0.028$.
a) As $b$ is decreased past $b=-1.009$, the fixed point becomes unstable via a supercritical Hopf bifurcation (the Hopf curve with label $n=1$ is crossed). As $b$ is decreased further, additional limit-cycles are created via supercritical bifurcations as Hopf curves with label $n=3,5,\ldots$ are crossed. b) The fixed point is unstable for $b > 1$. All bifurcations are supercritical.}
\label{fig:sinx}
\end{figure} 
%
%
As a first example, we consider $f(x)=\sin(x)$ as the nonlinearity in (\ref{model}). In this case, there are no quadratic terms in the Taylor-expansion, $f''(0) = 0$. Therefore, according to Corollary~\ref{cases_1_2}, the Hopf bifurcation is always supercritical because $f'(0) f'''(0) < 0$.
This is confirmed by the numerical results shown in Fig.~\ref{fig:sinx}, which are obtained using DDE-BIFTOOL \cite{dde-biftool}.  Displayed are the average amplitude of the periodic solutions versus the bifurcation parameter $b$, which is proportional to the feedback gain. 
Figure~\ref{fig:sinx} also demonstrates that all bifurcations are supercritical not only for $b>0$ but as well for $b=\gamma f'(0)<0$, thereby confirming that supercriticality is independent of the sign of $\gamma$.

Let us discuss in more detail the case of positive effective slopes $b$ shown in Fig.~\ref{fig:sinx}b. 
As $b$ is increased past $b=1$ the fixed point becomes unstable and a stable limit cycle is created. A further increase of $b$ results in the creation of additional limit cycles through supercritical bifurcations at the critical values of $b$. These additional limit cycles are stable within the center-manifold, but unstable with respect to the whole phase space, since the center-manifold itself has repelling normal directions. As these unstable limit cycles grow, additional bifurcations may occur that can change the stability property of the periodic solution. For example, in  Fig.~\ref{fig:sinx}b, the periodic-solution branch emanating from $b \sim 1.03$ is unstable for small amplitudes, as predicted by our theory, but is stabilized through a torus-bifurcation (Neimark-Sacker bifurcation of the corresponding map) at $b \sim 1.07$. In this context, it is important to emphasize that normal-form theory is valid sufficiently close to the fixed point and therefore makes statements only about small amplitude limit cycle solutions.

Finally, the sufficient condition for global stability provided by  Proposition \ref{stable_fp} yields the maximal range $|b|<1$, since $k_f=f'(0)=1$, consistent with supercritical Hopf-bifurcations.

\subsection{Nonlinearity without Cubic Terms}

\begin{figure}
   \centerline{\includegraphics[width=\columnwidth]{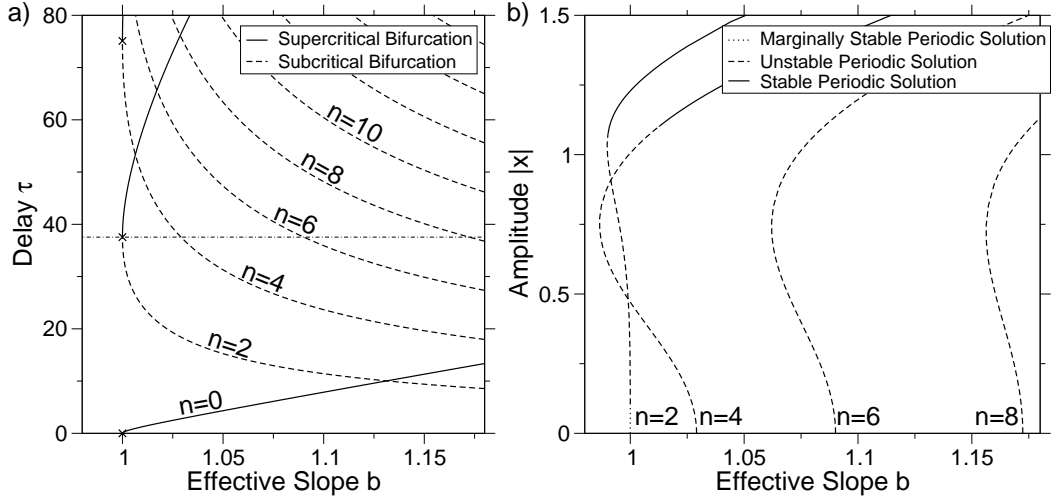}}
   \caption{a) Theoretical results: Shown are in the $b-\tau-$parameter plane the Hopf curves $(b_{C}^n(s),\tau_C^n(s))$, given by (\ref{bc}) and (\ref{tauc}) with $r=0.028$. The curves correspond to the location of Hopf bifurcations. Dashed lines indicate subcritical and solid lines supercritical bifurcations. b) Numerical results: The periodic solutions that arise are explored for $\tau=2 \pi/\sqrt{r} \approx 37.5$ (the dash dotted line in the left panel crossing Hopf curves with $n=2,4,6,8$). Except for the $n=2$ branch, which is a marginal case, the bifurcations are subcritical. }
\label{fig:xsqr}
\end{figure} 

As a second example, we consider a nonlinearity without cubic terms in the Taylor-expansion
\be
f(x)=(x + \frac{1}{2} x^2 + x^3 ) \, e^{-x^2}.
\ee
According to Corollary~\ref{cases_1_2} the type of the Hopf bifurcation is determined by the sign of $C^n(\tau,r)$. Proposition \ref{properties} establishes this sign for all possible values of $\tau$ and $r$ under the condition that $|b|<2$. 

To illustrate, we show in Fig.~\ref{fig:xsqr}a the Hopf curves  in the $b-\tau-$parameter plane and indicate whether the Hopf bifurcation on these curves is predicted to be subcritical or supercritical. 

Since the bifurcation is supercritical for $\tau > n \pi / \sqrt{r}$ and  subcritical for $\tau < n \pi/\sqrt{r}$, it follows that the case $\tau = n \pi/\sqrt{r}$ has to be marginal.
 Marginal means that the periodic solution branch emerges vertically  when the amplitude of the limit cycles that are created at the bifurcation point is plotted versus the bifurcation parameter $b$.  This is shown in Fig.~\ref{fig:xsqr}b, where $\tau \simeq 2 \pi/\sqrt{r}$ and it is seen that the periodic solution branch created from the $n=2$ curve at $b=1$ emerges essentially vertically.  The other periodic solution branches shown  in Fig.~\ref{fig:xsqr}b are all found to be created via subcritical Hopf bifurcations, in agreement with the predictions of Corollary~\ref{cases_1_2} and Proposition \ref{properties} (compare Fig.~\ref{fig:xsqr}a and  Fig.~\ref{fig:xsqr}b).

\subsection{Experiment of Blakely et al.}
\label{subsec:laspaper}

As our third and final example we  consider the optoelectronic device described in Ref.~\cite{laspaper}. We first map the model given in \cite{laspaper} onto (\ref{model}) and then show that our theory correctly predicts the experimental results.

In \cite{laspaper}, the following model is derived for the optoelectronic device with bandpass feedback
\begin{equation}
\begin{split}
\tau_h \dot{P}(\tilde{t}) =& -\left( P(\tilde{t})-P_0 \right) + \tau_h \kappa \dot V(\tilde{t}), \\
\tau_l \dot V(\tilde{t}) =& - V(\tilde{t}) + \tilde{\gamma}~G\left[ P(\tilde{t}-T_D)\right], \\
G\left[ P(\tilde{t}-T_D)\right] =& P(\tilde{t}-T_D)\left\{ 1+\beta~\sin \left[ \alpha \, \left(P(\tilde{t}-T_D)-P_0\right) \right] \right\},
\label{laspaper_model}
\end{split}
\end{equation}
where $P$ denotes the measured optical power and $V$ the voltage in the electronic feedback loop. The parameter values are taken from Table~I in Ref.~\cite{laspaper}.
These equations can be cast in the form of (\ref{model}) by introducing the rescaled and dimensionless variables 
$t = \tilde{t} ( \tau_h^{-1}+ \tau_l^{-1})$,
$x = (P-P_0)/P_0$, and 
$y = \tau_h (\tau_h+ \tau_l)^{-1}  \left\{ (P-P_0) - \kappa (V-\Tilde{\gamma} G[0])\right\} P_0^{-1} $.
The nonlinear delay term $f$ of (\ref{model}) is defined as
$f(x(t-\tau)) = \tau_h(\tau_l + \tau_h)^{-1} \left\{ G[ P_0 (x(t-\tau) + 1)] - G[P_0] \right\} P_0^{-1}$
and is given by
\be
 f(x)
= \frac{\tau_h}{\tau_h+ \tau_l}
 \left( x+\beta~x~\sin(a \, x)+\beta~\sin(a \, x) \right)\label{jonsnonl}. 
\ee
The fixed parameters of the model are $\tau=29.8$,  $r=0.028$, $a=\alpha P_0=49.14$, $\tau_h=22$ ns, $\tau_l=0.66$ ns, and $\beta=0.8$. For the Taylor-series of $f$, we obtain  $f'(0) = 39.14$, $f''(0) = 76.33$, and $f'''(0)= - 92163$. The values of the bifurcation parameter $b$ attainable in the experiment are $b \in [0,3.28]$, where $b$ is defined through $b := \kappa \tilde{\gamma} f'(0)$.

To illustrate the results of our analysis, we determine the critical gain and the type of the Hopf bifurcation using the above model.
The boundary of the local stability region can be estimated using (\ref{bcapprox}). We obtain $b_C= 1.003$, which corresponds to a gain $\tilde{\gamma} = 5.34$ mV/mW, in agreement with the experimental result of $\tilde{\gamma}=5.1 \pm 0.5$ mV/mW (see as well Fig. 3 of \cite{laspaper}).
In the experiment, it was found that the bifurcation is supercritical. Since $f'(0) f'''(0) < 0$, it follows from Corollary~\ref{cases_3_4_5} that the bifurcation is supercritical if $b$ is negative (negative gain) and is supercritical for positive $b$ if $\tau>n \pi/\sqrt{r}$. However, a close inspection of the inset of Fig.~\ref{fig:stabbdry} reveals that, for the given parameters ($\tau= 29.8$ and $r = 0.028$), the steady state loses its stability as the parameterized curve with branch-label $n=2$ is crossed to the left of the extrema, i.e. $\tau=29.8<2 \pi /\sqrt{r}$. Consequently, their experiment does not fall into one of the simple cases, necessitating a numerical evaluation of $C^n(\tau,r)$, which reveals that $f'(0) f''(0) + f''(0)^2 C^n(\tau,r) <0$. Thus, the bifurcation is predicted to be supercritical in agreement with the experimental result.

\section{Discussion}
\label{sec:discussion}

In this paper, we take a first step in analyzing rigorously the dynamics of two-dimensional DDEs that model delay-systems with band-limited feedback. We provide a condition guaranteeing global stability of the steady state solution and use linear stability analysis to derive the boundary of the local stability region in parameter space. We show that, generically, the steady state will lose stability in a Hopf bifurcation. The use of center-manifold techniques allows us to give a criterion for the type of this bifurcation which holds for arbitrary nonlinear functions $f$. Some effort went into simplifying this criterion to make it easier to predict whether supercritical or subcritical behavior will occur in a given experiment.

As mentioned in the introduction, the limit where high-pass filtering is removed from the feedback, i.e. in the limit $r \rightarrow 0$ in (\ref{model}), the two-dimensional DDE studied in this paper reduces to the class of scalar DDEs that includes the Ikeda DDE. It is informative to compare our results to known facts about Ikeda-type DDEs, since this brings into focus the novel features that arise due to the presence of the high-pass filter. 

The first and obvious difference is that blocking feedback at zero frequency in band-pass systems prevents the occurrence of multiple coexisting steady state solutions that exist for the case of scalar DDEs. Consistent with this observation, we find that fold-bifurcations do not exist in band-pass systems.
For scalar as well as two-dimensional DDEs, Hopf bifurcations occur and give rise to limit cycles. However, there is a fundamental difference concerning the frequencies (or modes) of these oscillations. 
For Ikeda-type DDEs,  the limit cycle at onset is always the fundamental ``period-2'' mode, with a frequency $f \sim (4 T)^{-1}$ for small delays and a frequency $f \sim (2 T)^{-1}$ for large delays ($T \gg \tau_l$) \cite{Nardone,Erneux}. In contrast, high-pass filtering results in a stability boundary where the mode at threshold varies with the chosen delay. As an example, in delay systems with negative band-limited feedback ($b<0$), the $n=1$ mode is observed for small delays. This mode corresponds to the  ``period-2'' mode of Ikeda-type DDEs because it has a frequency $f \sim (4 T)^{-1}$ for $T \approx 0$ and a frequency $f \sim (2 T)^{-1}$ for $T=2 \pi \omega_{max}^{-1}$. As the delay is increased past the timescale associated with maximum transmission ($T=2 \pi \omega_{max}^{-1}$), modes with $n=3,5,\ldots$ ($f \sim n/(2 T)$) form the stability boundary.  Thus, AC-coupling introduces fundamental differences to Ikeda-type DDEs already at the level of steady-state bifurcations, and the effects of high-pass filtering are particularly pronounced for large delays.

Also in this paper, we show that double-Hopf bifurcation points exist for delay systems with band-limited feedback, whereas these cannot exist in first-order scalar DDEs with a single delay. However, since double-Hopf points are found in n$^{th}$-order scalar DDEs \cite{Buono}, which may be used to represent multiple-pole low-pass-filter feedback, their appearance should be seen as a result of the increase in degrees of freedom rather than as a result of high-pass filtering.

Let us conclude with some remarks concerning the double-Hopf bifurcation points. In experiments, the chance of choosing a fixed delay such that upon changing the gain the stability boundary will be crossed close to a double-Hopf point is negligible. However, the existence of double-Hopf points can provide insight concerning possible dynamics for feedback gains above threshold. As an example, double-Hopf interactions lead to limit cycles as well as tori \cite{guckenheimer}. Consistent with this observation, torus attractors were found numerically in \cite{laspaper} for the example in Sec.~\ref{subsec:laspaper}. Furthermore, chaotic dynamics exists near a generic double-Hopf bifurcation \cite{guckenheimer}.
It would therefore be interesting to study how the coefficients of the double-Hopf normal form are related to the parameters of (\ref{model}) and to determine whether there are restrictions on the flow. Restrictions means that the mapping of the  parameters of the DDE onto the normal-form coefficients does not allow certain combinations of the double-Hopf normal-form coefficients to occur and thereby forbids the qualitative dynamics associated with those combinations. That such restrictions might occur was shown for instance by Buono {\em et al.} for the case of n$^{th}$-order scalar DDEs \cite{Buono}.

\appendix

\section{Appendix: Proof of Proposition \ref{stable_fp}}
\label{proof-globstab}

We are interested in investigating the stability of the trivial solution  of the initial-value problem for (\ref{model}), i.e. an equation of the form
\be
\dot{\vec{x}}(t)=g(t,\vec{x}_t),
\label{RFDE}
\ee
where the operator  $g(t,\varphi), g: \Reals \times B_H \rightarrow \Reals^d$ is continuous and Lipschitz in $\varphi \in B_H$ and $g(t,0)=0$.

To prove Proposition \ref{stable_fp}, we use an extension  of the method of Liapunov-Krasovskii functionals due to Kolmanovskii and Nosov \cite{Kolmanovskii}.  With the Liapunov-Krasovskii approach, difficulties arise because it is necessary to construct a functional with negative-definite derivative.
 It is often easier to find Liapunov-functionals with nonpositive derivative. For this case stability  may still be proved.

To show stability of the trivial solution, two continuous functionals $V(t,\varphi)$ and $W(t,\varphi)$ are used. We require that the derivative of $W$ is integrally unbounded. More precisely, designate $B_H$ as the closed ball of radius $H$ in the Banach space $C=C([-\tau,0],\Reals^d)$
\be
B_H = \left\{ \varphi \in C, d(0,\varphi) \le H \right\}, \nonumber
\ee
where $d(x,y)=\| x-y \|$ denotes the distance.  The
 the derivative $\dot W$ is called \emph{integrally unbounded} in a set $S \subseteq B_H$ if, for any number $K>0$, there exists a number $T(K)>0$ and a continuous function $\xi(t)$ such that uniformly in $x_t=x(t+\theta) \in S$ for $t \ge t_0$
\be
\dot{W} \le \xi(t), \quad \int_{t}^{t+T(K)} \xi(s) ds \le -K. \nonumber
\ee
We now cite the two theorems needed for the proof:
\begin{thm}[Theorem 5.7, pp. 77, \cite{Kolmanovskii}]
\label{kolmanovskii-theorem}
It is necessary and sufficient for uniform asymptotic stability of the trivial solution of (\ref{RFDE}) that there exist the functionals $V(t,\varphi)$ and $W(t,\varphi)$ such that
\begin{itemize}
\item[\normalfont (1)] $\omega_1\left(|\varphi(0)|\right) \le V(t,\varphi) \le \omega_2\left( \| \varphi(\theta) \| \right), \; t \ge t_0, \; \varphi(\theta) \in B_H, \; t_0 \in \Reals$. Here,  $\omega_i, i=1,2$, are some scalar, continuous, nondecreasing functions, such that $\omega_i(0)=0$ and $\omega_i(s)>0$ for $s>0$.  
\item[\normalfont (2)] $\dot V \le \overline{\omega}_3(\vec{x}_t) \le 0$, where  $\overline{\omega}_3(\varphi)$ is a continuous functional defined on $B_H$;
\item[\normalfont (3)] $|W(t,\varphi)| \le L, \; t \ge t_0, \; \varphi(\theta) \in B_H, t_0 \in \Reals$;
\item[\normalfont (4)] for any $\mu \in (0,H)$ there exists a $\rho > 0$ such that the derivative $\dot W$ is integrally unbounded in the set $E(\mu,\rho) \subseteq C$, where 
\be
E(\mu,\rho) = \left\{ \varphi(\theta) \in B_H, d\left(\varphi(\theta), S( \overline{\omega}_3=0 ) \right)  \le \rho, \mu \le \| \varphi(\theta) \| \le H   \right\} \nonumber.
\ee
Here $ S( \overline{\omega}_3=0 ) = \left\{ \varphi \in B_H, \overline{\omega}_3(\varphi)=0 \right\}$.
\end{itemize}
In addition, the attraction domain of the trivial solution is the ball $B_K$, where $K<H$ and $\omega_2(K) \le \omega_1(H)$.
\end{thm}
Furthermore, the following theorem establishes conditions for global stability:
\begin{thm}[Theorem 5.8, pp. 79, \cite{Kolmanovskii}]
\label{Kolmanovskii-global}
Let there exist two functionals $V(t,\varphi)$ and $W(t,\varphi)$ satisfying the condition of Theorem \ref{kolmanovskii-theorem} in any $B_H$ and
\be
\omega_1\left( |\varphi(0)| \right) \le V(t,\varphi), \quad \omega_1(s) \rightarrow \infty, \quad s \rightarrow \infty \nonumber
\ee
Then the trivial solution of (\ref{RFDE}) is globally uniformly asymptotically stable.
\end{thm}
Using Theorem \ref{kolmanovskii-theorem} and Theorem \ref{Kolmanovskii-global} it is straightforward to establish our claim of global stability of the trivial solution as stated in Proposition \ref{stable_fp}.

\begin{pf*}{Proof}
Consider (\ref{model}) with rescaled time $t \rightarrow 2 t$ and rescaled variable $y \rightarrow y/2$
\begin{equation}
\begin{split}
\frac{d x}{d t}(t) &= - 2 x(t) + y(t) + 2 \gamma f[x(t-\tau/2)] \\
\frac{d y}{d t}(t) &= - 4 r x(t).
\end{split}
\label{model_proof}
\end{equation}
Clearly, if the trivial solution of (\ref{model_proof}) is globally stable, then so is the trivial solution of (\ref{model}). Rescaling was chosen above because it optimizes the bound on the stability region. 
Using  (\ref{model_proof}) and the Liapunov-Krasovskii functional
\be
V(t,x_t,y_t)=\frac{x(t)^2}{2}+\frac{y(t)^2}{8 r}+\gamma^2 \int_{t-\tau/2}^{t} f[x(s)]^2 ds,
\ee
we obtain
\be
\dot{V}(t,x_t,y_t) 
= - \left\{ x(t)^2 -  \gamma^2 f[x(t)]^2 \right\}  - \left\{  \gamma f[x(t-\tau/2)] - x(t) \right\}^2 \nonumber \\
\le 0.   \nonumber 
\ee
The inequality holds because of the assumptions of Proposition \ref{stable_fp}, i.e. $ k_f |x(t)| > |f[x(t)]|$ ($\forall x \ne 0$) and $|\gamma|<k_f^{-1}$. 

Condition (1) of Theorem \ref{kolmanovskii-theorem} can be satisfied with $\omega_1(s)=s^2/2 $ and $\omega_2(s)= [(4 r)^{-1}+ \tau/2 ] s^2$. 
Condition (2) is satisfied with  $\overline{\omega}_3(\vec{x}_t) = \dot{V}(t,x_t,y_t)$.
We note that the set $ S( \overline{\omega}_3=0 )$ consists of elements $\vec{x}_t=(x_t(\theta),y_t(\theta))^T \in C \left([-\tau/2,0],\Reals^2\right)$ such that $ f[x(t-\tau/2)] = x(t)=0$.
Define the second functional as
\be
W = -x(t) y(t).
\label{second_functional}
\ee
Functional (\ref{second_functional}) is bounded in any ball $B_H$ and thus condition (3) is satisfied.

Denote the zeros of $f$ by $x^*_j$ , i.e. $f(x^*_j)=0$, and let $k_L$ be a positive constant such that, for any of the zeros $x^*_j$ and all $x \in \Reals$, the Lipschitz condition $|f(x^*_j) -f(x)| \le k_L \, |x^*_j - x|$ holds.
Define the set $E(\mu,\rho)$ through  $\mu \le \| \vec{x}_t \| \le H$, $| x(t) | \le \rho$, and $|x^*_j - x(t-\tau/2)| \le \rho$. The last inequality implies that  $|f[x(t-\tau/2)]| \le k_L \, \rho $. In this set we have
\begin{equation}
\begin{split}
\dot W &= 2 x(t) y(t) - y(t)^2 -2 \gamma  f[x(t-\tau/2)] y(t) + 4 r x(t)^2  \\
&\le 2 \rho H - \mu^2 + 2 \frac{k_L}{k_f} \rho H +  \rho^2  \\
&\le - \mu^2/2,
\end{split}
\nonumber
\end{equation}
where $r \le 1/4$ and the assumptions of Prop. \ref{stable_fp} were used.
The last inequality follows because $\rho$ may be chosen such that $\rho^2+2 (1+k_L/k_f) H \rho < \mu^2/2$.
Thus, $W$ is integrally unbounded as demanded by condition (4). 

Asymptotic stability of the trivial solution of model (\ref{model_proof}) follows from Theorem \ref{kolmanovskii-theorem}. Furthermore, all conditions of Theorem \ref{kolmanovskii-theorem} are satisfied for any ball $B_H$ and $\omega_1(s) \rightarrow \infty$ as $s \rightarrow \infty$. Thus,  the steady state of model (\ref{model}) is globally uniformly asymptotically stable. 

\end{pf*}

\section{Proof of Proposition \ref{main_theorem}}
\label{proof_main_theorem}

Using the identities
\begin{xalignat}{2}
\cos(\Omega \tau) &= \frac{1}{b}, &
\sin(\Omega \tau) &=\sgn(\tau-\frac{n \pi}{\sqrt{r}}) \frac{\sqrt{b^2-1}}{b}, 
\nonumber \\
\cos(2 \Omega \tau) &= \frac{2-b^2}{b^2} , &
\sin(2 \Omega \tau)  &= 2\sgn(\tau-\frac{n \pi}{\sqrt{r}}) \frac{\sqrt{b^2-1}}{b^2}, \nonumber
\end{xalignat}
and 
\be
4 \Omega^2 - r  =- \Omega \, \sgn(\tau-\frac{n \pi}{\sqrt{r}}) \sqrt{b^2-1} + 3 \Omega^2, \nonumber
\ee
the terms in the denominator of $K_2$ can be simplified. Define
\be
D_1 := D_1' + i D_1''  = b \left[ b \tau \Omega^2 + (r + \Omega^2) e^{i \Omega \tau} \right] ,
\nonumber
\ee
then 
\be
  D_1'= (b^2 \Omega^2 \tau + r+\Omega^2) \quad \text{and} \quad D_1''=\sgn(\tau-\frac{n \pi}{\sqrt{r}})(r+\Omega^2) \sqrt{b^2-1}. 
\nonumber
\ee
Applying the identities to
\be
D_2  := D_2' + i D_2'' = \Omega^{-1} b^2  \left[ 2 i \Omega b + (4 \Omega^2 - 2 i \Omega - r) e^{2 i \Omega \tau} \right], 
\nonumber
\ee
one obtains
\begin{equation}
\begin{split}
 D_2'  &= \sgn(\tau-\frac{n \pi}{\sqrt{r}}) \sqrt{b^2-1} (b^2+2) - 3 \Omega (b^2-2)   \\
 D_2'' &= 2 \left( b^3-1+\sgn(\tau-\frac{n \pi}{\sqrt{r}}) 3 \Omega \sqrt{b^2-1} \right).
\end{split}
\nonumber
\end{equation}
Use of the definitions of $D_1$ and $D_2$ allows us to rewrite $K_2$ (see (\ref{K2})) as
\bea
K_2 &=&
\frac{1}{2} \Real\left(\frac{b \, b_3 \Omega^2}{ D_1'+i D_1''} \right) + 
 \Imag \left( \frac{b^3 \, (b_2)^2 \Omega^2}{ (D_1'+i D_1'')(D_2'+i D_2'')} \right) \nonumber\\ 
 &=&
\frac{D_1' \Omega^2}{2 \left| D_1 \right|^2} \left\{ b \, b_3 + (b_2)^2 \frac{- 2 b^3  }{ D_1' \left| D_2 \right|^2} (D_1' D_2'' + D_1'' D_2') \right\}.
\nonumber
\eea
Note that $D_1'>0$ and therefore the sign of $K_2$ is given by the expression in `\{\}'-brackets. The definitions of $b$, $b_2$, and $b_3$, i.e. (\ref{def_b}) and (\ref{def_b2_b3}) respectively, imply $b \cdot b_3 = \gamma_C^2 f'(0) f'''(0)$ and $(b_2)^2= \gamma_C^2 f''(0)^2$. Using this and the definition
\be
\label{def_C_proof}
N^n= D_1' D_2'' + D_1'' D_2' \quad \text{and} \quad
C^n= \frac{- 2 b^3 \;  N^n}{ D_1' \left| D_2 \right|^2} 
\ee
yields the statement of Proposition \ref{main_theorem}.

\section{Proof of Proposition \ref{properties}}
\label{proof_of_C}

It suffices to determine the sign of $N^n$ because
the sign of $C^n$ is given by the sign of the product $-b N^n$ (see \ref{def_C_proof}). 
For clarity, let us reproduce here  (\ref{Nfunc})
\bea
N_{\lessgtr}^n&=& (r + \Omega^2 ) (b^4 +2 b^3+b^2-4) + 2 \Omega s b^2 (b^3-1) \nonumber \\
&& +\sgn(\tau-\frac{n \pi}{\sqrt{r}}) 3 \Omega \sqrt{b^2-1} [ (4-b^2)(r+\Omega^2) + 2 \Omega s b^2 ] \label{Nf}.
\eea
In writing (\ref{Nf}), we use the parameterization variable $s=\Omega \tau > 0$ and introduce the notation $N_>^n$ ($N_<^n$) for the case $\tau>n \pi/ \sqrt{r}$ ( $\tau<n \pi/ \sqrt{r}$). Note that distinguishing $N_>^n$ and $N_<^n$ allows us to view $N_{\lessgtr}$ as function of $b$, $N_{\lessgtr}$=$N_{\lessgtr}^n(b,r)$. The main idea behind this proof is to rewrite the sum (\ref{Nf}) in such a way that every term of the sum can be shown to have the same sign.  Throughout the proof, use is made of the fact that $\Omega$ and $\tau$ are strictly positive real numbers and that $0<r \le 1/4$. We address in turn the four cases of Proposition~\ref{properties}.

\begin{enumerate}
\item \emph{Case $\tau> n \pi / \sqrt{r} $:}
\begin{itemize}
\item[A)]
\emph{Case $b \in (1,\infty)$}: We need to show that $C^n<0$, which is true if $N_>^n >0$. Rewrite $N_>^n$ as follows:
\bea
N_{>}^n&=& (r + \Omega^2 )(b+2)\left[(b-1)(b^2+b+2) + 3 \Omega \sqrt{b^2-1}(2-b) \right] \nonumber \\
&& + 2 \Omega s b^2 (b^3-1) + 6 \Omega^2 s b^2 \sqrt{b^2-1}.
\eea
Clearly $N_>^n>0$ for $1 < b \le 2$. For $b>2$ it suffices to show that the first term in the sum is positive. However,
\bea
&&\left[(b-1)(b^2+b+2) + 3 \Omega \sqrt{b^2-1}(2-b) \right] \nonumber \\
&& \qquad \qquad  > (b-2)(b^2+b+2 - 3 \Omega \sqrt{b^2-1})
\eea
and, using the fact that $\Omega< \sqrt{r} \le 1/2$, implies
\be
(b-2)(b^2+b+2 - 3 \Omega \sqrt{b^2-1})>(b-2)(b^2-\frac{1}{2} b+2) >0. \nonumber
\ee
%
\item[B)] \emph{Case $b \in (-\infty,-1)$}: 
To show that $N_>^n < 0$, and therefore $C^n < 0$, we rewrite $N_>$ as follows:
\bea
N_>^n &=&
-(r + \Omega^2 )(b + 2)(1 - b)(b^2 + b + 2) - 
  2 \Omega^2 \tau b^2 (1 - b^3) \nonumber \\
&& - 3 \Omega \sqrt{b^2 - 1} (r + \Omega^2)(4 - b^2) - 
  6 \Omega^3 \tau \sqrt{b^2 - 1} b^2 . \label{case2a}
\eea
For $-1 > b \ge -2$ all terms of the sum are less or equal zero and the second term is negative definite. To show that $N_>^n<0$ for $b < -2$ we make use of the identity
\be
r+\Omega^2 = 2 \Omega^2 + \sgn(\tau-\frac{n \pi}{\sqrt{r}}) \Omega \sqrt{b^2-1}. \label{rpw2}
\ee
Substituting  (\ref{rpw2}) into (\ref{case2a}) and rearranging we obtain
\bea
N_>^n &=&
b^5 \left[ \frac{6}{5}s - \frac{5+6 \Omega s}{4}  \frac{\sqrt{b^2-1}}{|b|} \right] 
- (2 s - 12 \Omega) b^2 
\nonumber \\
&&
+ \left[ \frac{4}{5} s b^5 + 5 \Omega b^2 + 24 \Omega^2 \sqrt{b^2-1} \right]
- b^4 \left[ \Omega + 2  \frac{\sqrt{b^2-1}}{|b|} \right]
\nonumber \\
&&
+ b^3 \left[ (b^2-4) \frac{1+6 \Omega s}{4}  \frac{\sqrt{b^2-1}}{|b|} \right]
-20 \Omega - 4 \Omega^2 \sqrt{b^2-1}.
\label{case2b}
\eea
Note that the inequalities
\be
s>\pi>3, \qquad \Omega < \frac{1}{2},\quad \text{ and} \quad \frac{\sqrt{3}}{2}<  \frac{\sqrt{b^2-1}}{|b|} < 1
\ee
hold for $\tau> n \pi / \sqrt{r} $ with $b<-2$. Thus,
\bea
 \frac{6}{5}s - \frac{5+6 \Omega s}{4}  \frac{\sqrt{b^2-1}}{|b|}  &>& \frac{9}{20} \pi - \frac{5}{4}  >0 , \\
 2 s - 12 \Omega &>& 2 (\pi-3) > 0 , \\ 
 \frac{4}{5} s b^5 + 5 \Omega b^2 + 24 \Omega^2 \sqrt{b^2-1} &<&  \frac{12}{5} b^5 + \frac{17}{2} b^2 < 0 ,\label{case2blst}
\eea
where (\ref{case2blst}) holds for $b<-2$, because 
\be
\frac{12}{5} b^5 + \frac{17}{2} b^2 = \frac{107}{80}b^5 + \frac{17}{16} b^2 (b^3+8) < 0. \nonumber
\ee
Hence, the first three terms of  (\ref{case2b}) are negative definite and this is also true for the remaining terms.  
\end{itemize}
\item \emph{Case $\tau= n \pi/\sqrt{r}$, $n$ even $\rightarrow b=1$ :} 
From (\ref{Nf}) it can be seen immediately that $C^n=N^n=0$.
\item \emph{Case $\tau< n \pi/\sqrt{r}$, $b \in (1,2]$ :} We need to show that $C^n>0$, which is equivalent to showing that $N_<^n < 0$. Rewrite $N_<^n$ as
\bea
N_{<}^n&=& (r + \Omega^2 )(b+2) \left[ (b-1) (b^2 +b+2) - 3 \Omega (2-b) \sqrt{b^2-1} \right] \nonumber \\
&&
 - \frac{4}{3} s \Omega^2 b^2 \sqrt{b^2-1}
+ 2 \Omega s b^2 \left[-\frac{7}{3} \Omega \sqrt{b^2-1} + b^3 -1 \right]
\label{case3a}.
\eea
The expression in brackets in the first term of (\ref{case3a}) can be bounded from above by using the inequalities 
$s>\frac{3 \pi}{2}$, $\sqrt{b^2-1}< \Omega < \sqrt{b^2+2r-1}$, $r \le \frac{1}{4}$, and $(r+\Omega^2)< b^2 $. It is found that
\bea
&&(b-1) (b^2 +b+2) - 3 \Omega (2-b) \sqrt{b^2-1} \nonumber \\
&& \qquad < 4 (b-1) (b^2-\frac{b}{2}-1) = 4 (b-1) (b-\frac{\sqrt{17}+1}{4}) (b+\frac{\sqrt{17}-1}{4}). \nonumber
\eea
Using the above bound, we show that the sum of the first two terms of (\ref{case3a}) is negative. Consider the first two terms of (\ref{case3a}) 
\bea
&&(r + \Omega^2 )(b+2) \left[ (b-1) (b^2 +b+2) - 3 \Omega (2-b) \sqrt{b^2-1} \right] \nonumber \\
&& \qquad \qquad - \frac{4}{3} s \Omega^2  b^2 \sqrt{b^2-1}  \nonumber \\
&<& 4 (r + \Omega^2 )(b+2)(b-1) (b^2-\frac{b}{2}-1) - 2 \pi (b^2-1) b^2 \sqrt{b^2-1} \label{case3b}  \\
&<& 0. \nonumber
\eea
Here, the last inequality follows because both terms in  (\ref{case3b}) are nonpositive  for $b \in (1,(\sqrt{17}+1)/4]$ and it holds that
\bea
&&  4 (r + \Omega^2 )(b+2)(b-1) (b^2-\frac{b}{2}-1) - 2 \pi (b^2-1) b^2 \sqrt{b^2-1} \nonumber \\
&<& b^2 (b-1) \left[ 4 (b+2) (b^2-\frac{b}{2}-1) - 2 \pi (b+1) \sqrt{b^2-1} \right]  
\nonumber
\eea
for $b \in ((\sqrt{17}+1)/4,2]$, where it is straightforward to check that
\bea
 4 (b+2) (b^2-\frac{b}{2}-1) - 2 \pi (b+1) \sqrt{b^2-1} < 0. \nonumber
\eea
The remaining term of  (\ref{case3a}) is also negative on $b \in (1,2]$ because
\bea
 2 \Omega s b^2 \left[-\frac{7}{3} \Omega \sqrt{b^2-1} + b^3 -1 \right] 
&<&  2 \Omega s b^2 \left[ -\frac{7}{3} (b^2-1)+b^3-1 \right] \nonumber \\
&=&  2 \Omega s b^2  (b-1)(b-2)(b+\frac{2}{3}) \nonumber \\
&\le& 0. \nonumber
\eea

\item \emph{Case $\tau \le n \pi/ \sqrt{r}$, $b \in [-2,-1]$ :}
By rewriting $N_<^n$ as
\bea
N_{<}^n&=& - (r + \Omega^2 )(b+2)(1-b)(b^2 +b+2) - 2 \Omega s b^2 (1-b^3) \nonumber \\
&& - 3 \Omega \sqrt{b^2-1} (4-b^2)(r+\Omega^2) -  6 \Omega^2 s b^2 \sqrt{b^2-1}
\eea
it is immediately obvious that $N_<^n<0$ and therefore $C^n<0$.
\end{enumerate}

\ack

This work is supported by the U.S. Army Research Office under Grant DAAD 19-02-1-0223.

%
%


\begin{thebibliography}{99}
\bibitem{bbounds} The idea is to show that all double-Hopf points occur for $|b|<2$. Define $x_1^n=n \pi + \pi/3$, $x_2^n=n \pi - \pi/3$ with $n=0, 1, 2,\ldots$. Then $b_C^n(x_1^n)=b_C^{n+2}(x_2^{n+2})$ and $|b_C^n(x_1^n)|=2$, but, $\tau_C^{n}(x_1^n)>\tau_C^{n+2}(x_2^{n+2})$. The only part of the stability boundary that is not limited by the extrema-points ($|b|=1$) and double-Hopf points is the part of the boundary belonging to branch $n=1$ for which $b \rightarrow -\infty$ as $\tau \rightarrow 0$.
\end{thebibliography}
\end{document}